\documentclass[aps,reprint,twocolumn]{revtex4-1}
\pdfoutput=1
\usepackage{amsmath}
\usepackage{graphicx,color}
\usepackage{booktabs}
\usepackage{datetime}
\definecolor{navyblue!80!black}{rgb}{0.0,0.0,1}
\usepackage{xcolor}
\usepackage{epstopdf}
\usepackage{wrapfig}
\usepackage{ragged2e}
\usepackage{epsfig}
\usepackage{slashed}
\usepackage{parskip}
\usepackage{siunitx}
\newcommand{\mystar}{{\fontfamily{lmr}\selectfont$\star$}}

\begin{document}
%\title{Grain-grain interactions in streaming non-Maxwellian plasma}
\title{Plasma-grains interaction mediated by streaming non-Maxwellian ions}
%\title{Interplay of grain size and separation on inter-grain interaction in streaming non-Maxwellian plasma}

%\author{Sita Sundar}
\author{Sita Sundar$^1$ and Zhandos A. Moldabekov$^{2, 3}$}

%\affiliation{Institut f\"ur Theoretische Physik und Astrophysik, Christian-Albrechts-Universit{\"a}t zu Kiel, Leibnizstrasse 15, Kiel 24098, Germany}
\affiliation{$^1$Department of Aerospace Engineering, Indian Institute of Technology Madras, Chennai - 600036, India\\
$^2$~Institute for Experimental and Theoretical Physics, Al-Farabi Kazakh National University, 71 Al-Farabi str., 050040 Almaty, Kazakhstan}
%\affiliation{a b c}
\affiliation{$^3$Institute of Applied Sciences and IT, 40-48 Shashkin Str., 050038 Almaty, Kazakhstan}

\begin{abstract}
A comprehensive parametric study of plasma-grain interaction for non-Maxwellian streaming ions in steady-state employing particle-in-cell simulations is delineated. 
%, including shadowing force and nonlinear plasma response. 
Instead of considering the inter-grain interaction potential to be the linear sum of isolated grain potentials, we incorporate the numerical advancement developed fully  for grain shielding by including nonlinear contributions from  the plasma and shadowing effect.
  The forces acting on grains versus 
  inter-grain distance, streaming velocity of the ions, and 
  %as well as 
  impact of trapped ions density (number)  are characterized for non-Maxwellian ions 
   in the presence of charge-exchange collisions. 
   %Additionally, the trapped and untrapped ions density (number) is investigated. 
It is found that the nonlinear plasma response considerably modifies the plasma-grains interaction. Unlike the stationary plasma case, for two identical grains separated by a distance in the presence of streaming ions, the electrostatic force is neither repulsive for all grain separations nor equivalent to the force due to one isolated grain.  Inadequacy of  the linear response formalism in dealing with the systems having very large grain charges is also discussed. 
The smallest inter-grain separation for which the role of the shadow effect can be ignored is reported. 
\end{abstract}
\maketitle
%upto here 
\section{Introduction}
%Dusty (complex) plasma is a research domain of distinguished significance
 Dusty (complex) plasma is characterized by the presence of highly-charged dust particulates with sizes ranging from tens of nanometers to hundreds of microns in addition to  ions,  neutrals, and highly mobile electrons~\cite{Melzer:WVV2008}.  These charged particulates impart  features to the complex plasma physics leading to self-organization of dust particles,  propagation and instabilities of low-frequency dust particle waves~\cite{Ishihara:POP1997,barkan:POP1995,Kaw:POP1998, PRE_Moldabekov, Pintu:PRL2008,Amita:PRE2018}, formation of voids in experiments under micro-gravity conditions~\cite{Kretschmer:PRE2005}, 
ion-focusing and formation of dust induced wakes~\citep{Sundar:POP2017, Ludwig:EPJD2018, Sundar:PRE2018, Sundar:PST2019,Morfill:RMP2009,Kodanova}.

Wakefield drastically modifies interaction between charged particles \cite{Ludwig:EPJD2018, Moldabekov_CPP15, Moldabekov_CPP16}.
Dynamics of a grain in streaming plasma and wake formation is a problem of fundamental research interest and is one of the most discussed problems (e.g., see \cite{Lampe:POP2000,Ludwig:NJP2012}) due to its importance for a variety of applications ranging from fusion-related research~\cite{Winter:PPCF2004, Sharpe2002153, Ratyn:PPCF2008} to astrophysical topics~\citep{Halekas:SSR2011}, operation of gas discharges~\citep{Fortov:PRL2001},  understanding Langmuir probes~\citep{Hutch:POP2008}, and technological plasma applications~\cite{Hartmann:PRL2009,Hartmann:PRL2010}.  
Particularly, in radio-frequency (rf) discharge experiments, dust particulates develop near the sheath region where ions have non-zero streaming speeds due to the prevailing large-scale electric field.  

\begin{figure}[h!]
 \includegraphics[scale=0.45, trim = 0.0cm 10.5cm 9cm 2.5cm, clip =true, angle=0]{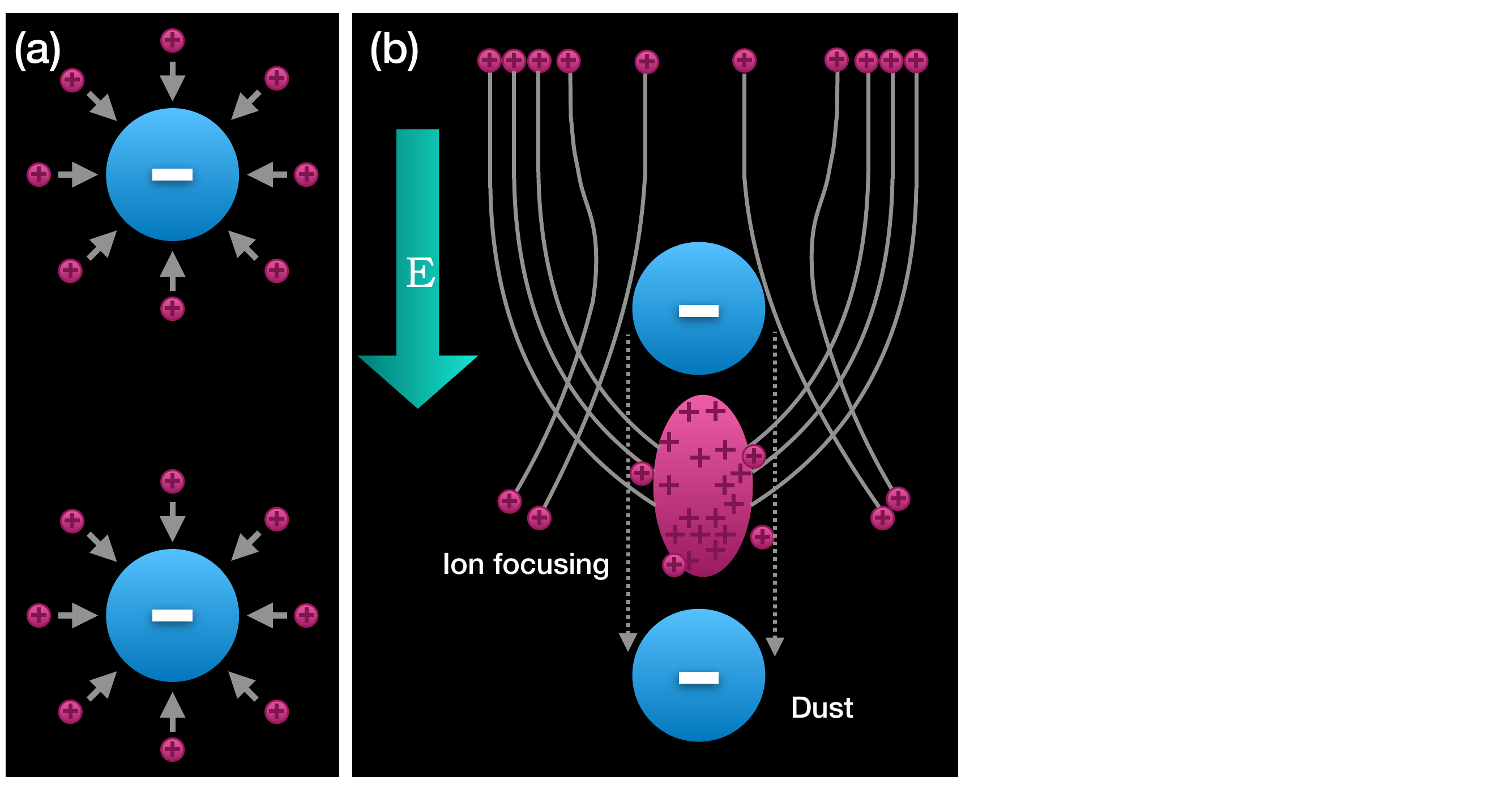}%
 \caption
 [Schematic illustrating  (a) a system of two stationary grains surrounded with ions isotropically  (b) an anisotropic system of two stationary grains in streaming ions in the presence of external driving field.]
{\protect\raggedright Schematic illustrating  (a) a system of two stationary grains surrounded with ions isotropically  (b) an anisotropic system of two stationary grains in streaming ions in the presence of external driving field. }
  % emphasizing the role of shadow force and non-linear response of plasmas
 \label{fig:figure1}
 \end{figure}

Impact of ion streaming on inter-grain interaction is illustrated here with a schematic in Fig.~\ref{fig:figure1}. In subplot (a) we have shown two identical stationary  grains isotropically screened by ions in stationary plasma. Subplot (b) exhibits two-grains  of same radii in streaming plasmas. It can be observed that the ions flowing towards the grain can be intercepted by the grains or  by an excess ion density focused downstream grain which could have otherwise bombarded the second grain (notice the dashed arrow towards downstream grain). Ions interact with the grains through coulomb forces as well as direct impact. Those ions which traverse nearby grain but encounter collision and have low enough energy get focused behind the grain. On the other hand, due to wakefield,  the potential profile in downstream direction has  positive regions \cite{Sundar:PRE2018, Ludwig:EPJD2018}. Therefore, grains passing upstream grain can get deflected by the cloud of focused ions.  From the subplots of Fig.~\ref{fig:figure1},  one can observe that the difference in the two cases emphasize the crucial role played by streaming of ions. 

 The grain-plasma system in the close proximity to the grain constitutes an open system (as illustrated in Fig.~\ref{fig:figure1})~\cite{Schweigert:PRE1996,Melzer:PRE1996,Schweigert:JETP1999}, which disapproves the calculation of interaction force directly from the derivative of electrostatic potential~\cite{Lampe:POP2015}. 
 One of the important  effects due to the finite size of the dust particles is a non-electrostatic force better named as `shadowing force' which has mechanical origin~\cite{Tsytovich:PPCF1996,Ignatov:PPR1996}. This shadowing force can be understood as the net ion-bombardment force exerted on a  grain towards every other grain  in its vicinity due to  the ions intercepted by each other. It is not a pairwise interaction force in the strict sense as it depends on the mutual alignment of the two or more grains and on their respective sizes.
 Work in this regard has been performed notably by Lampe {\it et al.}~\citep{Lampe:POP2015} wherein they  introduced the role of nonlinearity and `shadowing force' for simulating inter-grain interactions in stationary plasma and defied previous unjustified assumptions.
For the case of grains in stationary plasma,  Lampe {\it et al.}~\citep{Lampe:POP2015} showed that the role of trapped ions in manifesting any kind of attractive inter-grain interaction  is negligible and concluded that the electrostatic force between grains is always repulsive for the Maxwellian or shifted Maxwellian distribution of ions. 

In the presence of an electric field  driving the ion flow and ion-atom collisions (as is the case in the sheath region), the steady state ion velocity distribution is quite different from the Maxwellian and shifted Maxwellian distribution. The balance between electric field driven force and drag due to ion-neutral charge exchange collisions (the predominant collision) determine the flow velocity and the ion flow distribution function.  It has been shown using Monte-Carlo (MC) numerical scheme~\cite{Lampe:POP2012} that the resulting distribution for ions is non-Maxwellian.  Under constant ion-neutral charge-exchange collision frequency assumption, an exact solution for the distribution function reads~\cite{Sundar:POP2017,Lampe:POP2012,Ludwig:EPJD2018}, 
\begin{align}\label{eqn:drift}
f_z(u_z)&= \frac{1}{2M_\text{th}}\exp \left(\frac{1-2M_\text{th}u_z}{2M_\text{th}^2}\right) \times \nonumber\\
&\phantom{=}\left[1+\text{erf}\left(\frac{M_\text{th} u_z-1}{\sqrt{2}M_\text{th}}\right)\right],
\end{align}
here $u_z=v_z/v_\text{th}$ denotes the velocity  along the streaming direction in the units of the thermal velocity of neutrals, $v_\text{th}$,  and $M_\text{th}=v_d/v_\text{th}$ stands for the thermal Mach number with  $v_d$ being the drift velocity of ions.

For better understanding of the inter-grain interaction, 
  the non-Maxwellian ion distribution and nonlinear plasma response to the field of grains  have to be taken into account. Therefore, 
the purpose of the present work is to provide a consistent wide ranging numerical exploration of the forces acting on grains  and plasma-grain interaction using the particle-in-cell numerical approach. For two dust particles aligned along streaming velocity (see Fig. \ref{fig:figure1}), we simultaneously consider the impact of the following effects on grain-grain interactions : \\
\begin{itemize}
\item the non-Maxwellian velocity distribution of ions ;\\

\item the trapped ions in the vicinity of the grain;\\

\item the shadowing effect due to close location of grains;\\

\item the non-linear response of the plasmas.\\
\end{itemize}

The outline of the paper is as follows. In  Sec.~II and III, we introduce plasma parameters and simulation scheme utilized, respectively.  In Sec.~IV, we present the results for dust particle charges and plasma-grain interaction. Finally, in Sec. V. we discuss the perturbation of the density distribution of ions.

\section{Plasma parameters}
The investigation has been done at the following plasma parameters:
\begin{enumerate}
\item Electron temperature $T_e=3~{\rm eV}$, ion and neutral (Ar) temperature $T_i=T_n=0.03~{\rm eV}$,   $T_e/T_n=100$, plasma density $n_e=n_i=1\times 10^8~{\rm cm}^{-3}$, neutrals density (pressure) $n_n=5\times10^{14}~{\rm cm}^{-3}$ (${\rm P}= 2~{\rm Pa}$), the Debye length due to electrons $\lambda_{De}=\SI{1286.9}{\micro\metre}$, the Debye length due to ions $\lambda_{D}=\SI{128.69}{\micro\metre}$,  sound speed $c_s=\sqrt{T_e/m_i}=2.68\times10^5~{\rm cm/sec}$,  and  the ion charge number $Z_i=1$. 
\item The molecular species is Ar and is considered to be kept at a collision frequency which is related to the  collision cross-section  through $\nu_{in}=\sigma_{in}n_n v_{th}$ where $\sigma_{in}$ is the collision cross-section. The latter has been chosen to be equal to $\sigma_{in}=3.5\times 10^{-15}~{\rm cm^2}$, correspondingly the charge-exchange ion-atom collision frequency turns out to be $\nu_{in}=4.69\times10^4s^{-1}$. At considered plasma parameters, plasma frequency of ions turns out to be $\omega_{pi}=44.56~\nu_{in}$.
\item The dust particle has a cylindrical shape but with the diameter equal to the length.  The dust particle radius $a_d=\SI{5}{\rm \micro\metre} $, separation between dust particles $d/\lambda_D$ is varied from $2.02$ to $ 10.1$ with the step $1.01$ (for few cases with higher resolution) and $2.02$ (for most of the numerical simulations). Note that the grain radius is much smaller than the screening length, i.e.,  $\lambda_{De}/a_d=257.382$ and $\lambda_{D}/a_d=25.74$.
\item Ion streaming velocity considered herein encompasses the subsonic, sonic as well as supersonic regimes. The thermal Mach number $M_{\rm th}=0, 5, 10, 15$ is in the range from 0 to 15. The corresponding Mach number defined by the sound speed $M=M_{\rm th} \sqrt{T_n/T_e}$ varies in the range from 0 to 1.5. 
\end{enumerate}

\section{Numerical Details} 

We consider  a homogeneous plasma with ion flow driven by uniform ambient electric field, $E_0$ (as indicated in Fig.~\ref{fig:figure1}). Balance of the electric field and ion-neutral charge-exchange collisions, $\nu_{in}$, determine the ambient velocity distribution of ions, Eq.~(\ref{eqn:drift}).
The driving electric field is related to the charge-exchange collision frequency through the relation  $qE_{\rm drift}=m_i\nu_{in} v_d$, where $v_d$ is the drift speed of the ions, $m_i$ and $q=Z_ie$ denotes the mass and charge of the ion, respectively.

Numerical simulation has been performed with the two-dimensional (r,z) cylindrical  Particle-In-Cell (PIC) code `DUSTrz'~\cite{Lampe:POP2015}, where the grains are kept stationary. In `DUSTrz', the grain-grain separation and size could be varied and the dynamics of plasma is studied by following the motion of ions. Ions are PIC super-particles and electrons are taken to be thermal, i.e., the  electron density is given by  Boltzmann distribution $\Tilde{n}_e=n_{e} \exp (e\phi /T_e)$.
We have followed cgs unit for all physical parameters except temperatures which are in $eV$. 
%\cite ignatov of needed to describe homogenous electric field and boltzmann electron simultaneously

The equation to delineate the dynamics of the ions in the presence of self-consistent electric fields and driving electric field is given by
\begin{equation}\label{eq:motion}
    m_i\frac{d\mathbf{v}}{dt}=q\left[\mathbf{E}_{\rm pl}+\mathbf{E}_{\rm grain}\right],
\end{equation}
here  $\mathbf{E_{\rm pl}}$ is the field due to plasma (both electrons and ions) and $\mathbf{E_{\rm grain}}$ is the field due to dust grains.  The ions crossing the boundary of the simulation box are replaced by an ion chosen randomly in accordance with the ambient distribution. This way, the code is capable of incorporating a chosen  distribution.  We use non-Maxwellian distribution given by Eq.(\ref{eqn:drift}). Note that to avoid double counting the driving field $\mathbf{E_{\rm drift}}$ is not included explicitly in Eq.~(\ref{eq:motion}).

Simulation region is deliberately chosen to be very large ($10\lambda_D$ along radial direction and $40\lambda_D$ along the z-direction) to mitigate boundary effects. At the boundaries, electrostatic potential is set to zero.  Plasma space charge density along with grain charge density constitutes the source term for Poisson's equation.  Poisson's equation was solved in the given cylindrical simulation region at every ten time-steps. To facilitate the calculation of Poisson's equation,  grains are considered to have uniform charge density 
at all points on the grain surface  and any variation in the potential on the grain surface has been ignored. Moreover, one needs to resolve the ion dynamics in the vicinity of the grain and hence, the time-step has been chosen to be small $dt<\left(a_d/c_s\right)$ accordingly.
%Poisson equation  was solved at every ten time-steps. 

For computation of the dust particle charge, forces acting on dust particles for the non-Maxwellian ion distribution, the system is evolved  self-consistently for 3 000 000 time-steps and averaging is done over every 40000 time-steps. The weight used for simulation particles is of the order of unity and hence it is suitable for obtaining sufficient statistics of the particle motion.
%statistical fluctuations are a signature of real physics.

The present model is advanced in the sense that it does not consider the total potential for a system of two dust grains in streaming ions simply as linear sum of the potential due to two Debye spheres rather takes into account the effect of trapped ions and shadowing force  simultaneously incorporating the nonlinear response of plasmas in the grain-plasma system. 

\textit{Electrical force} acting on each dust particle  has been calculated as $Q_{1(2)}\left(\mathbf{E_{2(1)}}(z_{1(2)})+\mathbf{E_{pl}}(z_{1(2)})\right)$, where $\mathbf{E_{1(2)}}$ is the field due to grain 1 (2), $\mathbf{E_{pl}}$ is the field due to plasma. The dust particles are located along z axis (which is parallel to the drift velocity, see Fig.~\ref{fig:figure1}), where $z_1=0$ and $z_2=d$ (with $d$ being dust particles separation distance). In addition to the force due to grain charge and plasma, there is an additional force incorporated in the code in the same electrostatic force. This additional force which is also called ion drag, is due to the momentum deposition on the grains by ions traversing in the close proximity through Coulombic interactions. Moreover, electrostatic force here also takes care of the ions which get accelerated while passing in the neighborhood of the grain and fall into the electrostatic potential of the grain.

 \begin{figure}[h!]
 %\vspace{-25.5cm}
 \includegraphics[scale=1.0, trim = 0.cm 0.cm 0.cm 0.cm, clip =true, angle=0]{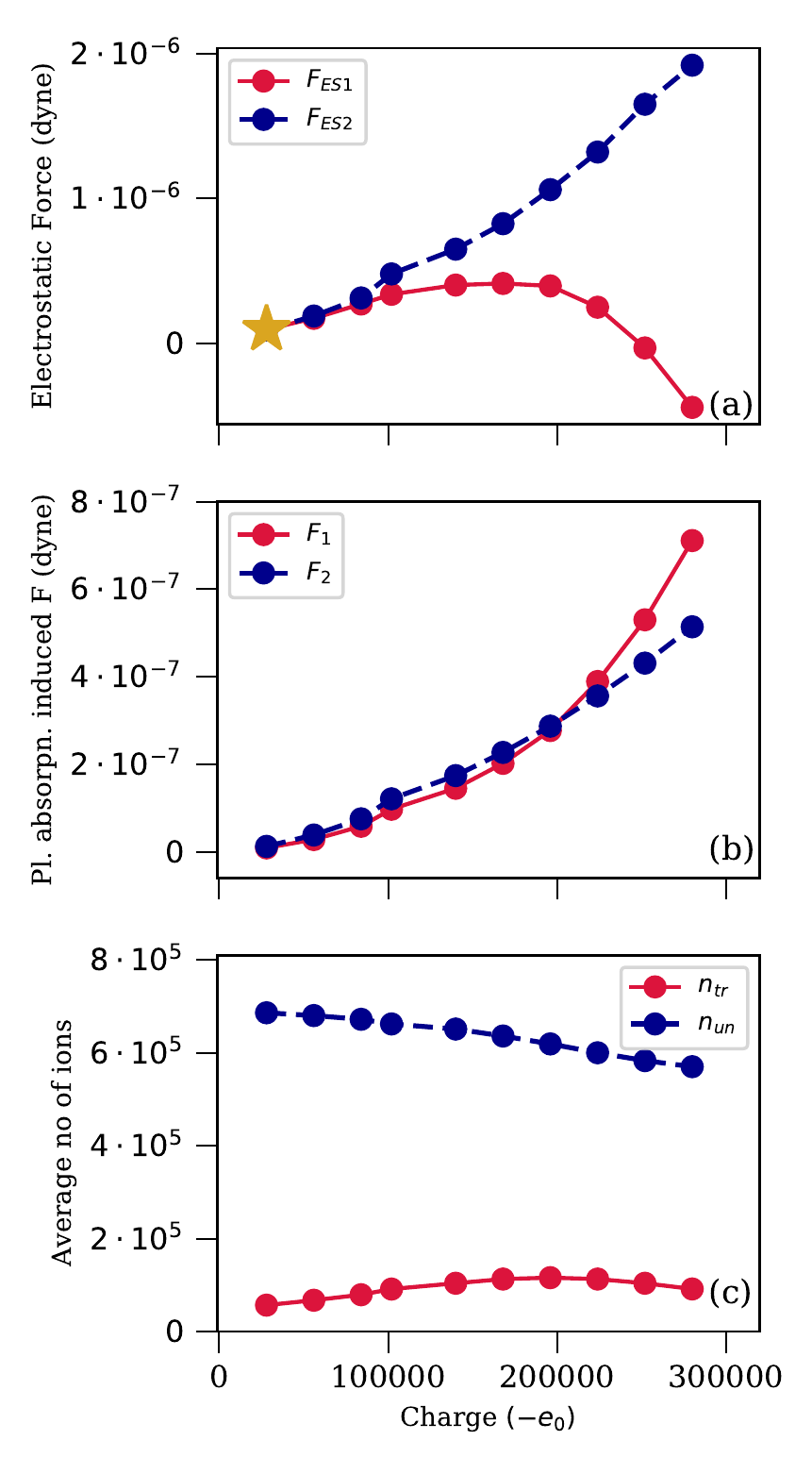}
 \caption{Inter-grain (a) Electrostatic forces, (b) Plasma absorption induced forces, and (c) number of trapped and untrapped ions as a function of  charges assigned to the the grains for $M_{th}=10$. 
 The charge on unit of electron is denoted by $e_0$ here. The inter-grain distance is fixed to $d/\lambda_D=6.06$.
 }
% because it is shadowed by the upstream particle.
 \label{fig:figure2}
 \end{figure}

\textit{Plasma absorption induced force} is the net rate of z-momenta deposited on the grain and is  computed 
 by  depositing the momenta on to a grain whenever an ion collides with a grain. It includes the momentum transfer due to the ions whose trajectory got intercepted by the other grain as well as the ions passing nearby grain whose trajectory is focused on the other grain surface. Due to symmetry, we calculate the momentum deposition along flow direction only. In the equilibrium plasma ($M=0$), the plasma absorption induced force is referred to as the shadow force, as one dust particle shadows the plasma flux on the 
 surface of the second  dust particle. In the case of single dust particle or very large  separation between dust particles, shadow force disappears.  However, for the case of streaming plasmas ($M\neq 0$), there is ion drag force which has one component due to scattered ions and the second component due to absorption of ions. Former is included into the mentioned electrostatic force and latter contributes to what we call here the plasma absorption induced force.  For two dust particles located close enough in streaming plasmas,  the  shadow force and plasma absorption related ion drag force can not be separated numerically from each other the way it is done theoretically. 

We  also computed the trapped and untrapped ions density (number).  According to standard Classical Mechanics, an ion with energy less than the maximum of the effective potential energy profile~\cite{Goree:PRL1992} is considered trapped. However, in the present work, an approximation in calculating the trapped ions has been utilized.
An ion is considered a trapped ion in the present simulation whenever its total energy is negative.  Number of untrapped ions is counted by integrating the difference between  the average ion density and untrapped (with positive total charge) ions density.

\section{Dust particle charges and forces}
 \begin{figure*}
  %[h!]
 %\vspace{-25.5cm}
\includegraphics[width=1\textwidth]{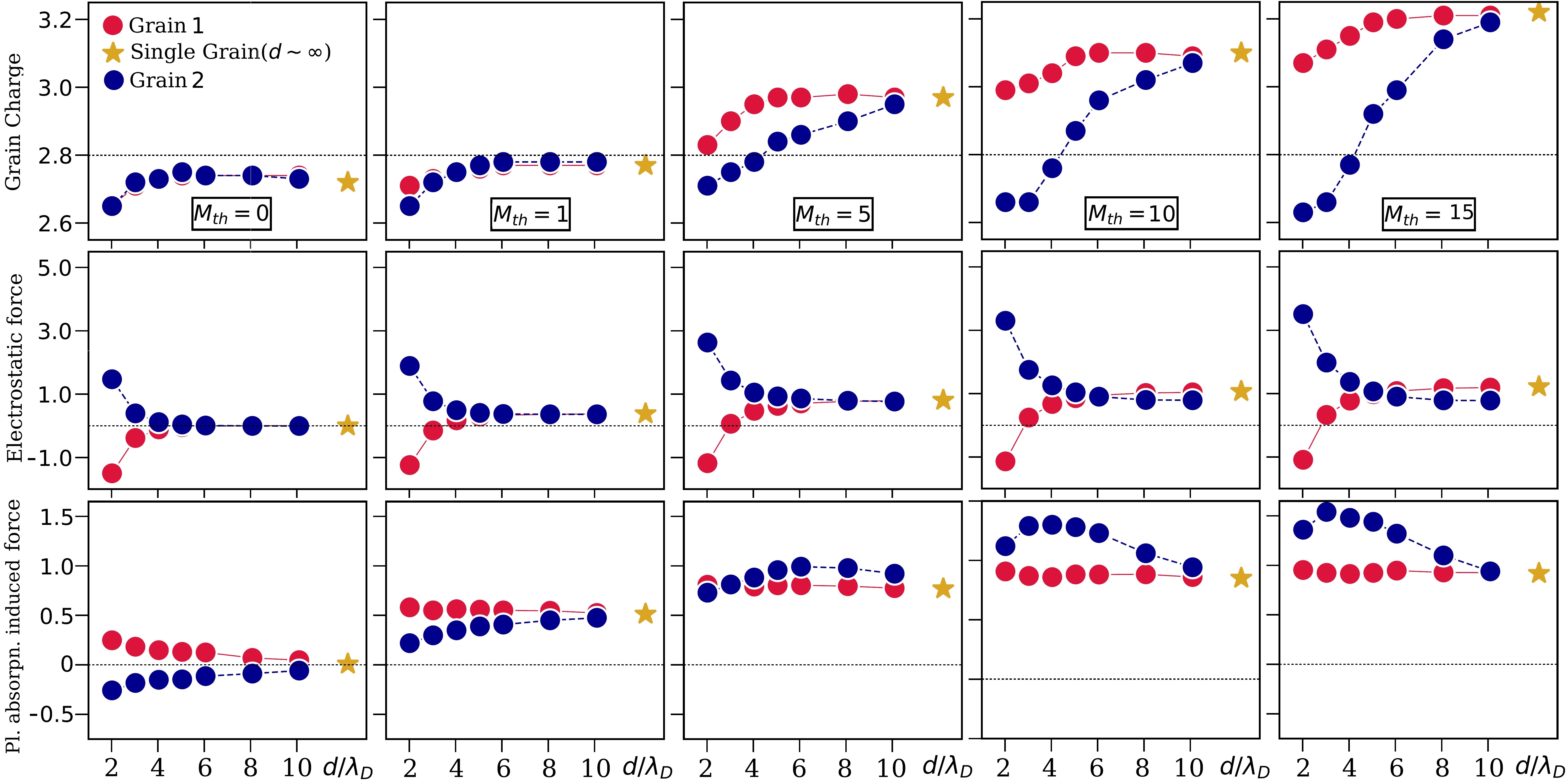}
 \caption{Grain charge [in units of $-e_0\times 10^4$, top row], Electrostatic force [in units of $dyne\times 10^{-7}$, middle row], and Plasma induced absorption force [in units of $dyne\times 10^{-8}$, bottom row] as a function of separation distance between the particles for  $M_{th}=0$, $M_{th}=1$,   $M_{th}=5$,   $M_{th}=10$, and  $M_{th}=15$.  The golden colored  {\mystar} %{\textcolor{yellow}{\mystar}} 
  symbol represents the corresponding data for single grain case ($d\sim\infty$).
 The black dashed horizontal line shown in the  subplot (top row) display the initial grain charge of $28000e_0$ where $e_0$ is the charge on one electron. The black dashed lines in the middle and bottom rows indicate the line at the value zero. }
% because it is shadowed by the upstream particle.
 \label{fig:figure3}
 \end{figure*}
 
 \begin{figure}%[h!]
 %\hspace{-3.5cm}
\includegraphics[width=0.41\textwidth]{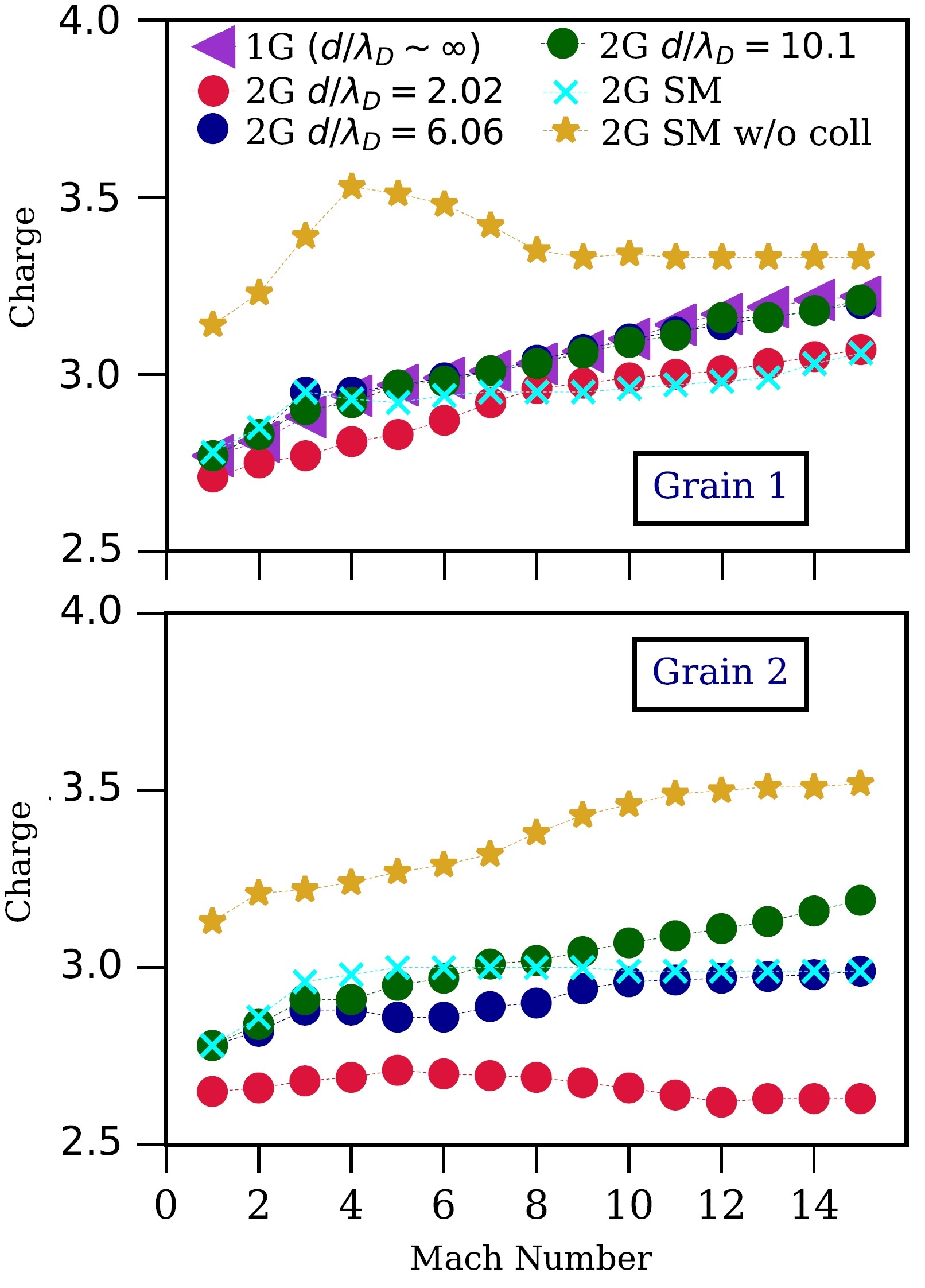}
\caption{Plot exhibiting the grain charge [in units of $-e_0\times 10^4$] versus streaming speed for (a) upstream for versus intergrain separations $d/\lambda_D=2.02,6.06,10.1$ along with single grain case  ($d/\lambda_D\sim\infty$) and (b) downstream grain for  inter-grain separations $d/\lambda_D=2.02,6.06,10.1$. Additionally, the data for the shifted Maxwellian distribution  with collision (SM) and  without collision (SM w/o coll.) are presented for $d/\lambda_D=6.06$. The plasma conditions are the same as in Fig.~\ref{fig:figure3}.}
\label{fig:figure7}
\end{figure} 
 
 \begin{figure*}%[h!]
 %\vspace{-25.5cm}
\includegraphics[width=1\textwidth]{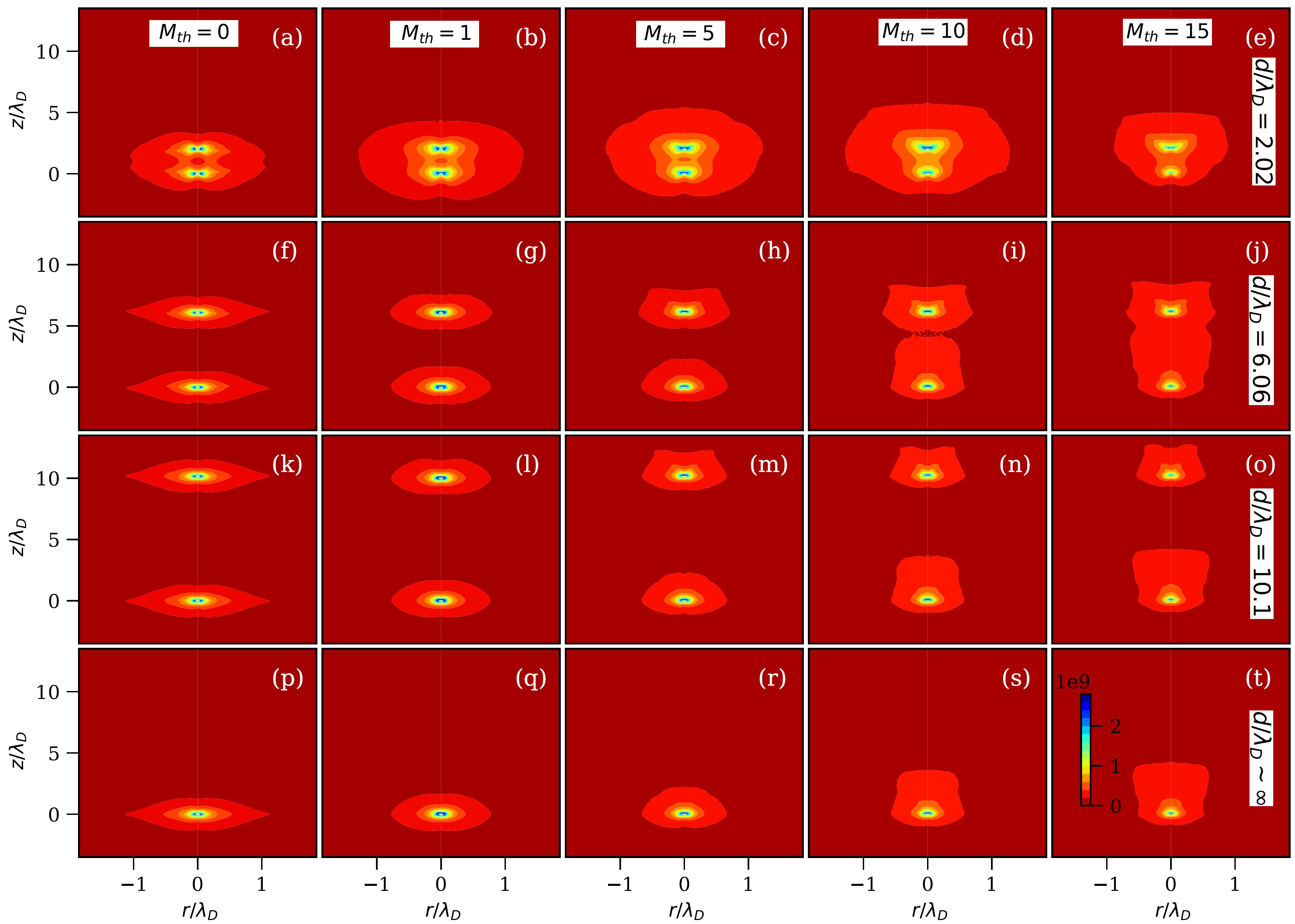}
 \caption{Spatial profiles of the induced ion density for  $M_{th}=0,1,5,10,15$ (column wise from left to right) for inter-grain separations $d/\lambda_D=2.02$ in subplots (a)-(e) (the top-most row), $d/\lambda_D=6.06$ in subplots(f)-(j) (second row from top), $d/\lambda_D=10.1$ in subplots (k)-(o) (third row from top) respectively for two grains,  and  in subplots (p)-(t) (the bottom-most row) for single-grain, all with non-Maxwellian drift distribution.
 %with inter-grain separation
%$d/\lambda_D=6.06$.
The plasma conditions are same as in Fig~\ref{fig:figure3}.}
% Grain charge as a function of separation distance between the particles for (a) $M_{th}=0$,  (b) $M_{th}=5$, (c)  $M_{th}=10$, and (d) $M_{th}=15$.  The plasma conditions are 2 Pascal Ar  with  $1 \times 10^{8} cm^{-3}$  ambient ion density, and 3.0 eV electron temperature. The black dashed horizontal line shown in the four subplots display the initial grain charge.}
% because it is shadowed by the upstream particle.
 \label{fig:figure4}
 \end{figure*}

% Complex plasmas characterized with large number of the parameters which can be varied. In addition to the plasma temperature and density, the dust particles charge, size, temperature and density can be varied. We ignore effects related to the latter two 
% by considering only two immobile grains. The approximation of immobile grains is justified due to drastic difference between the characteristic time scales of dust particles and that of ions as well as electrons. 
Considering different values of the dust particle charge allows to illustrate the manifestation
of the shadowing effect and non-linear plasma response.  
 We have assigned linearly increasing charges on the grain and have measured the forces acting upon grains and number of the trapped and untrapped ions. 
The dust particle charge, being linearly proportional to the grain size,  was computed self-consistently. 
  To inquire plasma-grain interaction for different values of the dust particle charge, the dust particle size was varied from  
  $\SI{5}{\rm \micro\metre}$ to $\SI{50}{\rm \micro\metre}$. %(\textcolor{red}{please check !}). 
Dependence of inter-grain force on  the charge assigned is shown in  Fig.~\ref{fig:figure2} for the fixed inter-grain distance of $d/\lambda_D=6.06$. From  the subplot (a) of Fig.~\ref{fig:figure2}, we clearly see that with the increase of the dust particle charge from $10^4$ (denoted by a star) to $10^5$ the forces (directed downstream) acting on the dust particles increase approximately linearly. Further increase in the dust particles charge clearly shows strong non-linear dependence of the force on the dust particle charge at $Z_d>10^5$.  
 In this strong non-linear regime, the electrostatic force acting on the downstream particle (denoted as $F_{ES2}$) continues to increase while the upstream particle (denoted as $F_{ES1}$) exhibits a non-monotonic behavior. It first increases with grain charge till it reaches its maxima where it saturates (at $a_d\simeq \SI{25}{\rm \micro\metre}$) %\textcolor{red}{please check !})
 followed by a decreasing trend eventually changing its sign upon further increase in the grain charge (from $a_d>\SI{35}{\rm \micro\metre}$ onwards).%  \textcolor{red}{please check !}).
 %continues to increase and the electrostatic force acting on the upstream particle (denoted as $F_{ES1}$) first saturates,  then decreases and finally  changes its sign (direction) with further increase in charge. 

  The reason behind non-monotonic behavior of the upstream grain can be the negative ion
drag force \cite{ PhysRevLett.100.055002, Momot} or Coulomb repulsion due to downstream particle. The negative ion
drag force appears due to the formation of  a negative charge region behind the grain (in downstream direction) resulting from strong absorption of ions on the surface of the dust particles. For equilibrium plasmas, Vladimirov et al  \cite{PhysRevLett.100.055002} reported the negative ion
drag force for single grain case considering an absorbing  grain with small size ($a_d/\lambda_D\ll 1$) in high  pressure  plasma (with the ion collection by grain characterized by continuum limit   $l_i/a_d \lesssim 1$). 
 In our numerical simulations,  we have $a_d/\lambda_D\simeq 0.27$ and $l_i/a_d \ll 1$. Clearly, despite lower pressure, the continuum limit for the ion collection by grain is also realized (due to large grain size). However, the electrostatic force acting on  downstream particle (with the same size) does not change its sign. Therefore, we conclude that the strong Coulomb repulsion between grains with large charge ($Z_d>2.5\times 10^5$) overcomes the ion drag force acting upon upstream particle manifesting in the change of the sign of the electrostatic force.
 
 %it is not obvious here that the s
 
 In  the subplot (b) of Fig.~\ref{fig:figure2}, the plasma absorption induced force is presented. With the increase in the dust particle charge, the plasma absorption induced force increases for considered values of the dust particle charge. Except the region around the grain charge value at which the electrostatic force acting on the upstream particle changes its sign, the   absorption induced force is approximately smaller by one order of magnitude.  In  the subplot (c) of Fig.~\ref{fig:figure2}, the number of trapped and untrapped ions is shown. The number of trapped ions exhibit a very small variation  and number of untrapped ions  decreases insignificantly  with increase in the dust particle charge. This may seem somewhat counter-intuitive at first glance, but easily can be  understood by recalling that the excess plasma (ions and electrons) density around dust particles is strictly controlled by the plasma quasi-neutrality condition \cite{Moldabekov_PST}, better understanding can be gained by looking at the pattern of the plasma distribution around dust particles. The latter is discussed in Sec.~VI. 
 
 The described dependence of the forces on the dust particle charge clearly illustrates its significance for the  variety of the possible phenomena. 
In the presence of streaming ions, two grains with the same size  can collect different charges depending on the inter-grain separation. We also notice the non-reciprocal behavior of forces for the two grains. Unlike the case of two identical stationary grains in stationary ions~\cite{Lampe:POP2015}, for the streaming ions the force on one grain exerted due to the other is no longer the same as the force exerted due to the field of an isolated grain exerting on the bare charge of the other. 
Flow velocity of ions, inter-grain separation, and non-linearity of plasma in the vicinity of the grain has  a crucial role in determining the steady state charge of the grains. 
Usually, in experiments the dust particle size is in the range $1$ - $\SI{10}{\rm \micro\metre}$. Therefore, we focus on the dust particles in this range and present a detailed analysis of the dust particles with size $a_d= \SI{5}{\rm \micro\metre}$ in our subsequent discussion in this paper. 
%further we consider in more detail the case with the dust particles size $a_d= \SI{5}{\rm \micro\metre}$.

To understand the inter-grain  and plasma-grain interactions,  in Fig.~\ref{fig:figure3} we have shown the grain charges and the forces acting on grains for various values of the streaming velocity and  inter-grain separation. The top most row  shows the grain charge (in units of electron charge) versus inter-grain separation distance (in units of $\lambda_D$)  for  ions with $M_\text{th}=0, 1, 5, 10, 15$ (from left to right) respectively.
The golden colored {\mystar} symbol in all the subplots of Fig.~\ref{fig:figure3} denotes the single-grain data. The  single grain data can also be understood as the case where two grains are completely isolated from each other (i.e. $d/\lambda_D\sim\infty$). The values for two grain case asymptote towards single grain case when one moves the two grains farther apart  beyond $d/\lambda_D=10.1$.
It is important to point out here that in the presence of streaming ions, action is not equal to reaction ({\it{actio$\neq$ reactio}}) and one can not clearly define an effective pair-interaction. It motivated us to
%That is one of the reasons to 
explore the effect of different kind of forces  explicitly on grain-plasma dynamics. 

 For $M_\text{th}=0$ (left column - top row), the grain charges on both the grains are equal in the steady-state for all the inter-grain separation distances. 
At $M_\text{th}=1$, the second column from left, the charge of dust particles differs at $d/\lambda_D<4$ and almost same at larger separation distances.  
In the remaining three subplots in the top row, the charge on the upstream particle (grain 1) at $d/\lambda_D\leq 10$ is higher in magnitude than the downstream particle (grain 2) indicating that  the downstream particle charges less negatively.  This is due to the fact that with streaming ions, focusing of ions occur downstream  giving rise to smaller negative charge for the downstream grain.
Though  the  charge on grain 1 is more than grain 2  at $d/\lambda_D\leq 10$ for streaming ions ($M_{th}\gg 1$), nevertheless, the difference between the two grain charges asymptote towards zero at very large separations (which is equivalent to the single grain case) where one expects the shadowing and focusing effects to be negligible. At smaller separations, the difference between grain charges increases with streaming ion speeds. In general, the pattern exhibited by grain-charge with respect to inter-grain separation distance is  due to the synergistic role played by shadowing effect and plasma streaming. 

To better illustrate the grain charge-Mach number dependence, in Fig.~\ref{fig:figure7}, we have presented %the plot illustrating 
  the grain charge versus streaming ion speed for (a) upstream grain as well as single grain and (b) downstream grain for varying inter-grain separation. The general trend suggests that the grain charge increases with streaming speed except for very small inter-grain separation $d/\lambda_D=2.02$ where the downstream grain exhibits a non-monotonic trend. 
  The reason for such a non-monotonic behavior can be attributed to a significant overlap of the shielding clouds [see Sec.~V].
   The upstream grain charge exhibits a monotonically increasing behavior with streaming speed albeit with no significant variation with inter-grain separation beyond $d/\lambda_D=6.06$. For completeness, in Fig.~\ref{fig:figure7}, we compare the data for $d/\lambda_D=6.06$ with that of obtained using the shifted Maxwellian distribution  with collisions as well as without collisions. For thermal Mach numbers $M_{th}>4$, the data computed using the shifted Maxwellian distribution with collisions shows lower values of the upstream grain charge in comparison to the non-Maxwellian case. Regarding the downstream grain charge, the non-Maxwellian  character of the distribution leads to the lower charge values at $2<M_{th}<10$. The important result is that the neglect of the ion-neutral collisions leads to a significant overestimation of the charge of both upstream and downstream dust particles and, hence, to an inadequate description of the electrostatic and the plasma absorption induced forces. In fact, the ion-neutral collisions lead to the increase of the trapped ions around grains. More detailed discussion of this effect is given in Sec. V.

Electrostatic force versus grain separation is shown in the second (middle) row of  Fig.~\ref{fig:figure3}.
For $M_{th}=0$, the force acting on the upstream grain 
is opposite to the force acting on the downstream grain. 
%{\textcolor{blue}{$M_{th}=1$ regime manifests counter-intuitive physical features for  small inter-grain distances. It shows a non-monotonic behavior of Electrostatic force versus inter-grain separations. }}
Similarly, for  $M_{th}\neq 0$ and $d/\lambda_D=2.02 $, the electrostatic force on grain 1 is asymmetrically opposite to grain 2.  In these cases the Coulombic repulsion between dust particles is stronger than the ion drag force. The total force acting on the system of two dust particles is zero in the case  $M_{th}=0$ and non-zero in the case $M_{th}\neq 0$.
As anticipated, electrostatic force is significant at smaller inter-grain separations and is weaker at larger separations. At $M_{th}\neq 0$ and $d/\lambda_D\geq 4$, the electrostatic force acting on both particles is positive and directed downstream.
For $M_{th}=10$ and $M_{th}=15$, at distances $d/\lambda_D \geq 6$, the electrostatic force acting on upstream particle (grain 1) becomes greater than that  acting on downstream particle (grain 2).
It means  at larger grain separations $d\geq 6.06\lambda_D$ electrostatic force is trying to push the upstream particle closer to the downstream particle.

In general, for $0\leq M_{th} \leq 15$, at separation distances  $d/\lambda_D\geq 5$ the electrostatic force is approximately constant and close to that of for a single grain case. A strong deviation of the electrostatic force from the single grain case appears at $d/\lambda_D< 5$ for all considered streaming velocities  ($0\leq M_{th} \leq 15$).

Moreover, the magnitude of electrostatic force is higher than  plasma absorption induced force (see bottom row) for all the grain separation distances. So, we can say that the electrostatic force is the dominant force as its magnitude is always larger than the shadow force. Note that the opposite situation may accrue at significantly larger grain charges (sizes) as it is illustrated in Fig.~\ref{fig:figure2}.
 \\
 
The plasma absorption induced force versus grain separation is shown in Fig.~\ref{fig:figure3} (bottom row). 
For $M_{th}=0$, the plasma absorption induced force is referred to as the shadow force. In this case, the shadow force acting on grain 1 is asymmetrically opposite to the force acting on grain 2. This means shadow force tends to attract dust particles to each other.
In a plasma with streaming ions, at relatively small distance between grains, the ion drag component due to plasma absorption on the surface of the dust particles can not be separated from the shadow force. However, at large enough distances between grains, the shadowing effect vanishes as the problem reduces to the two grains isolated from each other.  Indeed, at $d/\lambda_D=10.1$, for the considered case of equal sized dust particles, the plasma absorption induced force is almost same  for both upstream and downstream particles and has values close to the single grain case. In such a scenario, the plasma absorption induced force is the ion drag force component due to absorption of ions on the surface of the grain.
 
 For $M_{th} \neq 0$, the plasma absorption induced force has positive net value for the system of two dust particles, i.e. 
 both particles are dragged along streaming direction. Interestingly, at $M_{th}=1$ the upstream particle (grain 1) is pushed 
 stronger (due to plasma collection) than the downstream particle giving rise to the force that is similar to the attractive shadow force
 in equilibrium case ($M_{th}=0$). The same behavior is seen at small separation distance $d/\lambda_D=2$ at $M_{th}=5$.   
 This effect disappears with increase in the streaming velocity as one can see from the data for  $5\leq M_{th} \leq 15$ in Fig.~\ref{fig:figure3}, i.e. the downstream particle (grain 2) is pushed stronger than upstream particle (grain 1). Therefore, at these values of streaming speed the ion drag force due to plasma collection on the grain surface is dominant over shadowing effect related force.

Remarkably, at $M_{th}\geq 1$, the plasma absorption induced force acting on upstream grain is almost independent of the separation distance from downstream grain and has the value equal to the single isolated dust particle plasma absorption induced force. In contrast, the plasma absorption induced force acting on downstream grain shows a strong variation with the separation distance from upstream grain and approaches the value of the single isolated dust particle case only at very large distance ($d/\lambda=10.1$).   

%It has  positive value for upstream grain and equal negative value for downstream grain. 
%This repulsive effect at very small inter-grain sepa is due to the contribution from  the focusing of ions by one grain onto another. At $M_{th}\neq 0$, the plasma absorption induced force is always positive (meaning directed along streaming), implying that the ion drag force component due to absorption is always dominant over shadow force.  

\section{Ions perturbation by dust particles}
  %\begin{figure*}%[h!]
 %\hspace{-3.5cm}
% \includegraphics[scale=0.81, trim = 0cm 0cm 0cm 0cm, clip =true, angle=0]{Figure5.pdf}
% \caption{Variation of the (a) trapped and (b) untrapped ion density along the flow direction for the two grains. First (upstream) grain is at origin and the second (downstream) grain is placed at different locations $d/\lambda_D\sim 2,4,6,8,10$ for $M_{th}=5$ (left column),  $M_{th}=10$ (middle column), and  $M_{th}=15$ (right column). 
 %Here, $z_0$ is the unit length along the streaming direction.
 %}
 %\label{fig:figure5}
 %\end{figure*}
 
  \begin{figure*}%[h!]
 %\hspace{-3.5cm}
 \includegraphics[scale=0.81, trim = 0cm 0cm 0cm 0cm, clip =true, angle=0]{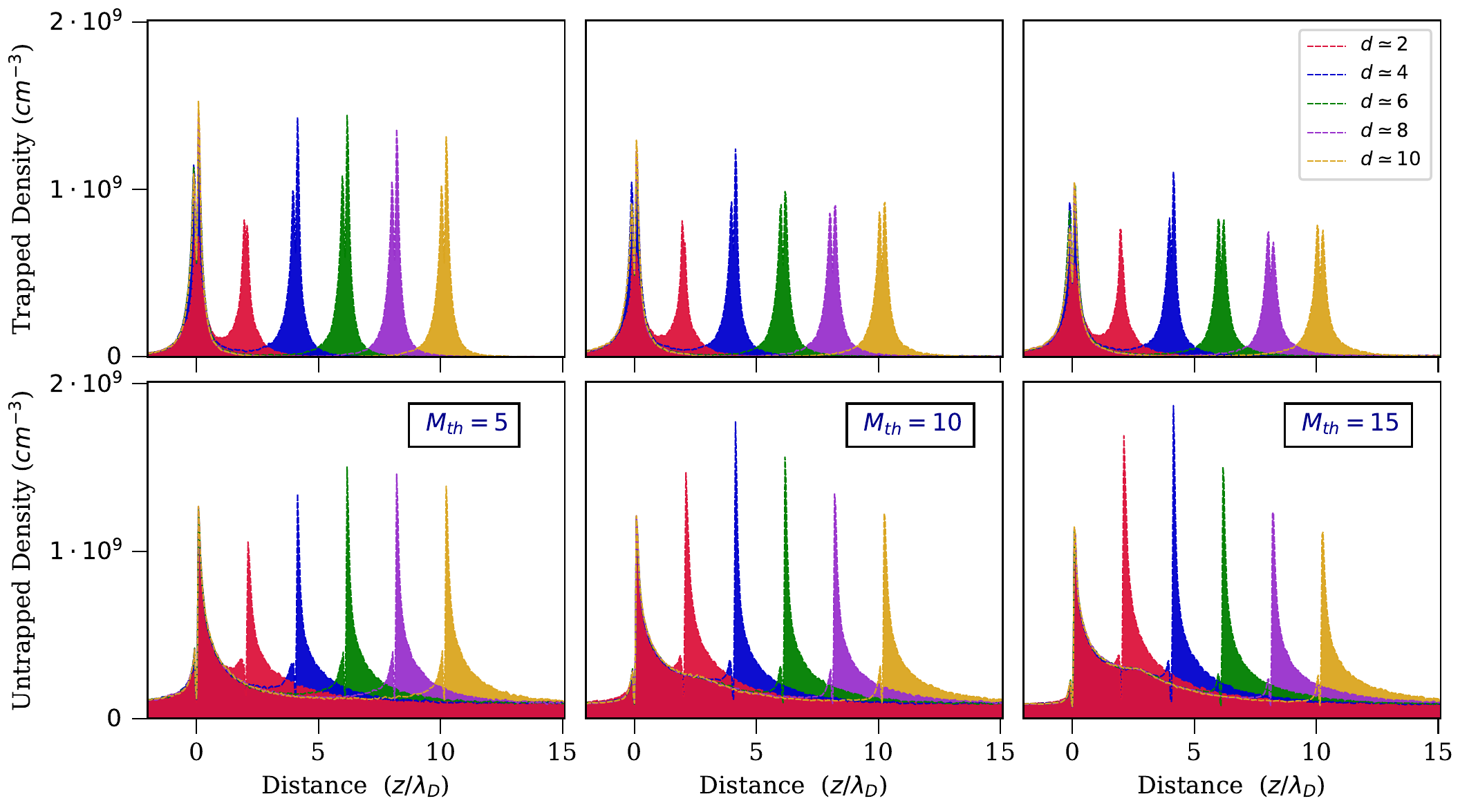}
 \caption{Variation of the (a) trapped (top row) and (b) untrapped (bottom row) ion density along the flow direction for the two grains. First (upstream) grain is at origin and the second (downstream) grain is placed at different locations $d/\lambda_D\sim 2,4,6,8,10$ for $M_{th}=5$ (left column),  $M_{th}=10$ (middle column), and  $M_{th}=15$ (right column). 
 %Here, $z_0$ is the unit length along the streaming direction.
 }
 \label{fig:figure5}
 \end{figure*}

 \begin{figure*}%[h!]
 %\hspace{-3.5cm}
\includegraphics[width=1\textwidth]{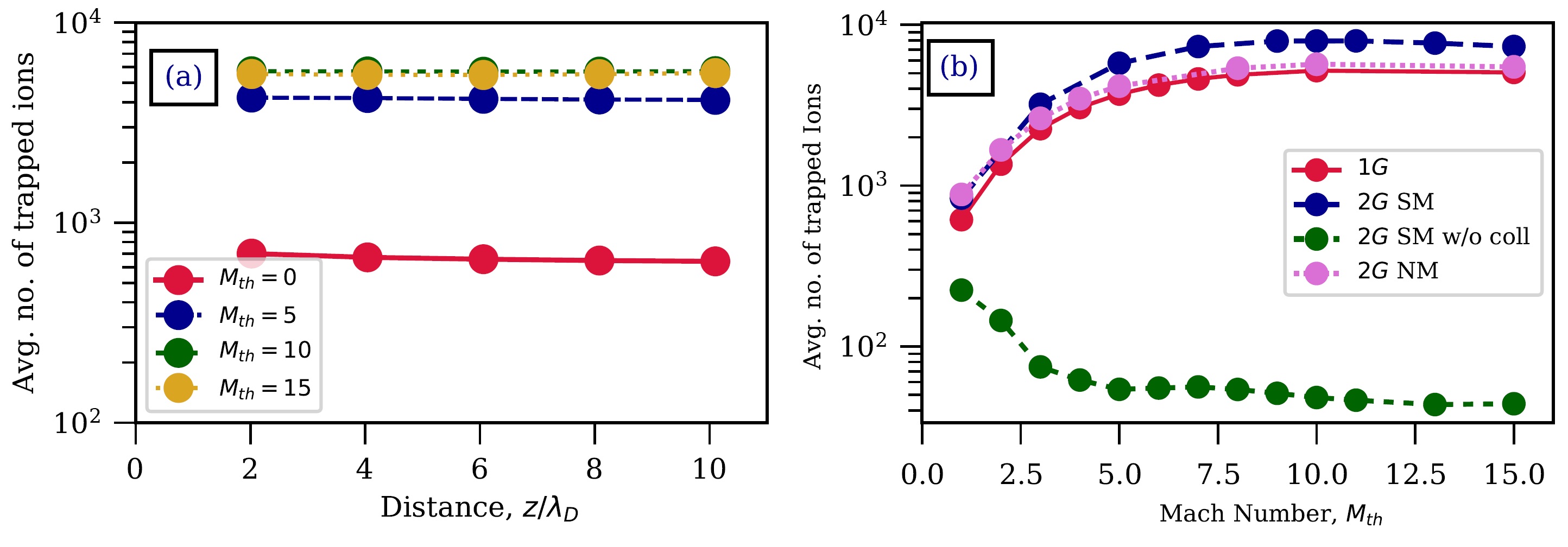}
\caption{Plot exhibiting average number of (a) trapped   ions as a function of separation distance between the particles for  $M_{th}=0$ (red circle - solid line), $M_{th}=5$ (blue circle - big dashed line), $M_{th}=10.0$ (green circle - small dashed line), and $M=15$ (yelow circle - dotted line). Subplot (b) illustrates the variation of trapped ion density versus streaming speed for Single Grain (red circle - solid line), two-grains at $d/\lambda_D=6$  for Shifted Maxwellian with collision (blue circle - big dashed line) and without collision (green circle - small dashed line) and two-grains with non-Maxwellian distributions (orchid circles - dotted line).  The plasma conditions are the same as in Fig.~\ref{fig:figure3}.}
\label{fig:figure6}
\end{figure*} 

Fig.~\ref{fig:figure4} shows the 2D ion induced density distribution around dust particles  for (a) $M_{th}=0$ (first column), (b) $M_{th}=1$ (second column), (c) $M_{th}=5$ (third column), (d) $M_{th}=10$ (fourth column), and $M_{th}=15$ (fifth column) (from left to right column-wise)  at the inter-grain separation $d/\lambda_D=2.02, 6.06, 10.1$ (top to bottom row-wise) and the case for single grain case (the bottom-most row). 
For the subsonic regimes $M_{th}\le5$, one can see well separated plasma polarization (symmetric in  both r and z) around the two grains except the case for very small inter-grain separations. At $d/\lambda_D=2.02$, the size of overlapping ion cloud increases with increase in flow up to $M_{th}=10$ and then decreases at very high streaming speed of $M_{th}=15$ (see subplots (a)-(e)). At larger inter-grain separation distances, as one increases the ion streaming speed, the plasma polarization around grain becomes strongly anisotropic (asymmetric in z) and the shielding around the two grains start to overlap each other (last two columns from left corresponding to $M_{th}=10$ and $M_{th}=15$ ). The single-grain case depicted in the bottom-most row (subplots (p)-(t)) manifests semblance with the two-grain case for large inter-grain separations.
 With increase in streaming ion speed, the size of the polarized ion cloud around dust particles increases as ions become scattered to large distances (smaller angles). 
  %\begin{figure}%[h!]
 %\hspace{-3.5cm}
%\includegraphics[scale=0.91, trim = 0cm 0cm 0cm 0cm, clip =true, angle=0]{Figure8}
%\caption{Plot exhibiting average number of (a) trapped and (b) untrapped ions as a function of ion flow speed between the particles for  single-grain non-Maxwellian (red circle - solid line), two-grain Shifted Maxwellian (blue circle - big dashed line), two-grain Shifted Maxwellian without collisions (green circle - small dashed line), and two-grain non-Maxwellian (orchid circle - dotted line).  The plasma conditions are same as in Fig.~\ref{fig:figure3}.}
%\label{fig:figure8}
%\end{figure} 
 This observation is supported by the results in Fig.~\ref{fig:figure5}, where ion density (trapped as well as untrapped) versus grain separation is shown. Trapped ions are the ions collected by the grain which have a total negative energy. Subplots (a) and (b) of Fig.~\ref{fig:figure5} show the trapped and untrapped density respectively for $M_{th}=5$, $10$, and $15$. The upstream dust particle (grain 1) is located at $z=0$ and the second dust particle (grain 2) is positioned at $z=d$. %{\textcolor{red}{
 From Fig.~\ref{fig:figure5}, one can observe that the downstream excess ion density  has monotonically decaying character without clearly distinct (separated) focused ion cloud. For a single dust particle, with  non-Maxwellian ion distribution, such pattern was previously reported in Ref. \cite{Moldabekov_PST} on the basis of the linear response approach. In the present work, we confirm this   pattern for the case of two dust particles beyond linear approximation. Overall, the trapped ion density does not show strong variation with change in the streaming velocity from $M_{th}=5$ to $M_{th}=15$. Nevertheless, one can note the localized nature of trapped ions around grains and spreading of untrapped ions with streaming ion velocity downstream grain. The untrapped ion density for streaming ions shows a sharp front and a long ion density tail. Two peaks (in red color one at $z=0$ and other at $z=2.02\lambda_D$ and similarly for other inter-grain separations) show the density around two grains. The peak of the first grain for all grain distances overlap at $z=0$.
 
Additionally, it can be seen from Fig.~\ref{fig:figure5}, the trapped ion density does not show much variation with change in inter-grain separation. This is in agreement with the computed value of grains charge, which also does not change drastically with increase in Mach number and the separation distance between grains. Note that the trapped ions, having high probability to fall on the surface of dust particle, is correlated with the effective charge of the dust particles.

For $M_{th}=5$, similar to the trapped ions case, the untrapped ion density does not exhibit a significant variation with increase in inter-grain separation. However, at higher values of streaming velocity,  $M_{th}=10$ and $M_{th}=15$, the untrapped ion density has its maximal values at $d/\lambda_D=4.04$ (see Fig.~\ref{fig:figure5}). One point which is noteworthy is that both trapped and untrapped ion density peaks around $d=6.06\lambda_D$ for subsonic flows. This peak shifts towards $d=4.04\lambda_D$ as we increase the flow strength. 

 In Fig.~\ref{fig:figure6} (a), for different values of Mach number, the dependence of the number of trapped  ions on the  inter-grain separation distance is shown. 
From this figure one can see that trapped  number of ions is approximately constant for all values of the inter-grain separation distance. With increase in the Mach number from $0$ to $10$, the number of trapped ions (i.e., ions with negative net energy) increases. Further increase  in streaming speed to $M_{th}=15$ leads to slightly smaller number of trapped ions in comparison with the case $M_{th}=10$. 
 
In Fig.~\ref{fig:figure6} (b), we have  shown the behavior of average number of trapped ions in the system with streaming ion speed for various ion distribution functions at $d/\lambda_D=6$. 
One can observe clearly that the number of trapped  ions is slightly larger for two-grains with non-Maxwellian distribution compared to single grain case with the same distribution. 
The trend exhibited by trapped ions with Mach number for the  non-Maxwellian case are increasing. This, at first, seems counter-intuitive, however, an analysis taking into account the impact of charge-exchange collisions reveal that it is not. To verify the cause of this peculiar behavior we have also shown alongside the plot for the cases with shifted Maxwellian with and without collision. We see that the two-grains without collisions exhibit decreasing trend with Mach number as expected. Inclusion of collisions lead to the increase in the number of trapped ions at $M_{th}\leq 10$. Indeed,  the ions passing in the vicinity of the dust particles can lose their energy  due to charge-exchange collision. 
Therefore, described non-monotonic dependence of the trapped ions number on Mach number can be understood as the competition between the effects of stronger influx of ions [losing energy due to collisions and getting trapped in the grain potential] and  higher escaping ability with increase in the Mach number.
We can say now with proof that collisions are the reason behind the increase in the average number of trapped ions. 

 The trapped and untrapped ions  shed light on ionic dynamics in the vicinity of the charged dust particles. 
 Besides, from the presented data in Secs. IV and V, we understand that the trapped ions  play a crucial role in grain charging and, therefore, in the resulting plasma-grain interaction force.  
 Moreover,  in linear regime as well as in linear response approach all ions are considered to be untrapped [i.e., an ion potential energy is much less than its kinetic energy], meaning that the fraction of the trapped ions can characterize nonlinearity of plasma polarisation in the vicinity of the grain. Due to quasi-neutrality condition in plasmas, the total negative dust particles charge is compensated by the polarized positive volume charge around grains. The data for considered values of the thermal Mach number and inter-grain separation distances reveals that in the case of non-Maxwellian distribution the fraction of the trapped ions in the polarized screening cloud around dust particles increase from $1~\%$ to $10~\%$ with increase of the thermal Mach number form $0$ up to $10$. In contrast, in the collisionless case, the fraction of the trapped ions decreases with the streaming speed and is always less than one percent. 
 %We found that at $2\leq d/\lambda_D\leq 10$ and $M_{th}>5$ the fraction of untrapped ions is about $10~\%$.
 
% This is in coherence with our observation of electrostatic and ion absorption induced force behavior which exhibit attractive behavior around $d=6.06\lambda_D$ for $M_{th}=5$ and the critical distance shifts towards smaller inter-grain separation as one increases the streaming speed of ions.

% Moreover,  in linear regime as well as in linear response approach all ions are considered to be untrapped, meaning that the fraction of the trapped ions can characterize nonlinearity of plasma polarisation in the vicinity of the grain.
 %We found that at $2\leq d/\lambda_D\leq 10$ and $M_{th}>5$ the fraction of untrapped ions is about $10~\%$.
 
\section{Conclusions}

We have presented here the study of grain charge and forces acting on grains at various parameters for non-Maxwellian ion flow distribution under typical experimental situations using particle-in-cell simulation scheme.
%We enhanced the accuracy of the model significantly by increasing the complexity of the model as we have taken into account the effects of several physical attributes simultaneously. 
As a result we observed that:
\begin{itemize}
%    \item  The force-charge dependence does not show linear character for $Z_d \geq  10^5$.
    \item With increase in the streaming speed,   the difference between the charge on the two grains increases while with increase in the inter-dust particle distance the difference between the charge on the two grains decreases.
    \item The plasma absorption induced force is not affected by the shadowing effect at $d/\lambda_D \geq  10$ and $M_{th}\leq 15$. At smaller inter-dust distances, the shadowing  effects  can not be  neglected if the plasma absorption induced force is considered. For upstream grain, at $M_{th}\geq 1$ and $d>2\lambda_D$, the plasma absorption induced
force is  approximately equal to an isolated
dust particle plasma induced absorption force.  For downstream grain, the plasma absorption induced
force  approaches the value of the single isolated grain case at $d/\lambda_D>10$.
    \item For the electrostatic force with streaming speeds in the range $0\leq M_{th} \leq 15$, the results obtained for the single grain case do not provide accurate description for $d/\lambda_D< 5$. In the case $M_{th}=10$ and $M_{th}=15$, unlike the case of stationary Maxwellian ions~\cite{Lampe:POP2015}, the total electrostatic force pushes the upstream and downstream grains towards each other at $d/\lambda_D\geq 6$.
    Therefore,  the electrostatic force for the identical grain-pair  in streaming ions with $10\leq M_{th}\leq 15$ is no longer repulsive for grain separations $d/\lambda_D\geq 6$. It gives an indication of the attractive force for the grain-pair and the possibility for the existence of a bound pair of grains as envisaged in the previous experiments~\cite{usachev2009formation}.
    \item The ion density perturbation due to two dust particles  has a  monotonically decreasing tail  in downstream direction. The number of trapped  ions [i.e., the ions with negative total energy] remains approximately constant  with increase in the inter-dust particle distance from $d/\lambda_D=2.02$ to $d/\lambda_D=10.1$. 
    In the collisional case the fraction of the trapped ions in the screening of the dust particles varies in the range from $1~\%$ to $10~\%$ and 
 increases with the streaming speed at $M_{th}\leq 10$, while in the collisionless case the opposite trend with streaming speed is realized with the fraction of trapped ions in the screening cloud always less than $1~\%$.
\end{itemize}

\section{Acknowledgements}
SS would like to thank Dr. M. Lampe for support in using the code `DUSTrz',  and acknowledges the support and hospitality of CAU Kiel Germany and IIT Madras India. 
%Stimulating discussions with Zhandos Moldabekov are gratefully acknowledged. 
%Author also appreciate the valuable scientific comments of Zhandos Moldabekov.
This work was supported by the DFG via SFB-TR24, Project A9 and DRDO project via project no. ASE1718144DRDOASAM.
Zh. Moldabekov thanks the funding from the Ministry 
of Education and Science of the Republic of Kazakhstan via the grant  BR05236730 
 ``Investigation of fundamental problems of Phys. Plasmas and plasma-like media'' (2019).  
%Numerical  simulations  were  performed at HPC cluster of Christian-Albrechts-Universit{\"a}t zu Kiel.

%\newpage

\bibliography{dusty_manus}
%\bibliography{mybib}{}
%\bibliographystyle{plain}
\include{dusty_manus.bbl}

\end{document}